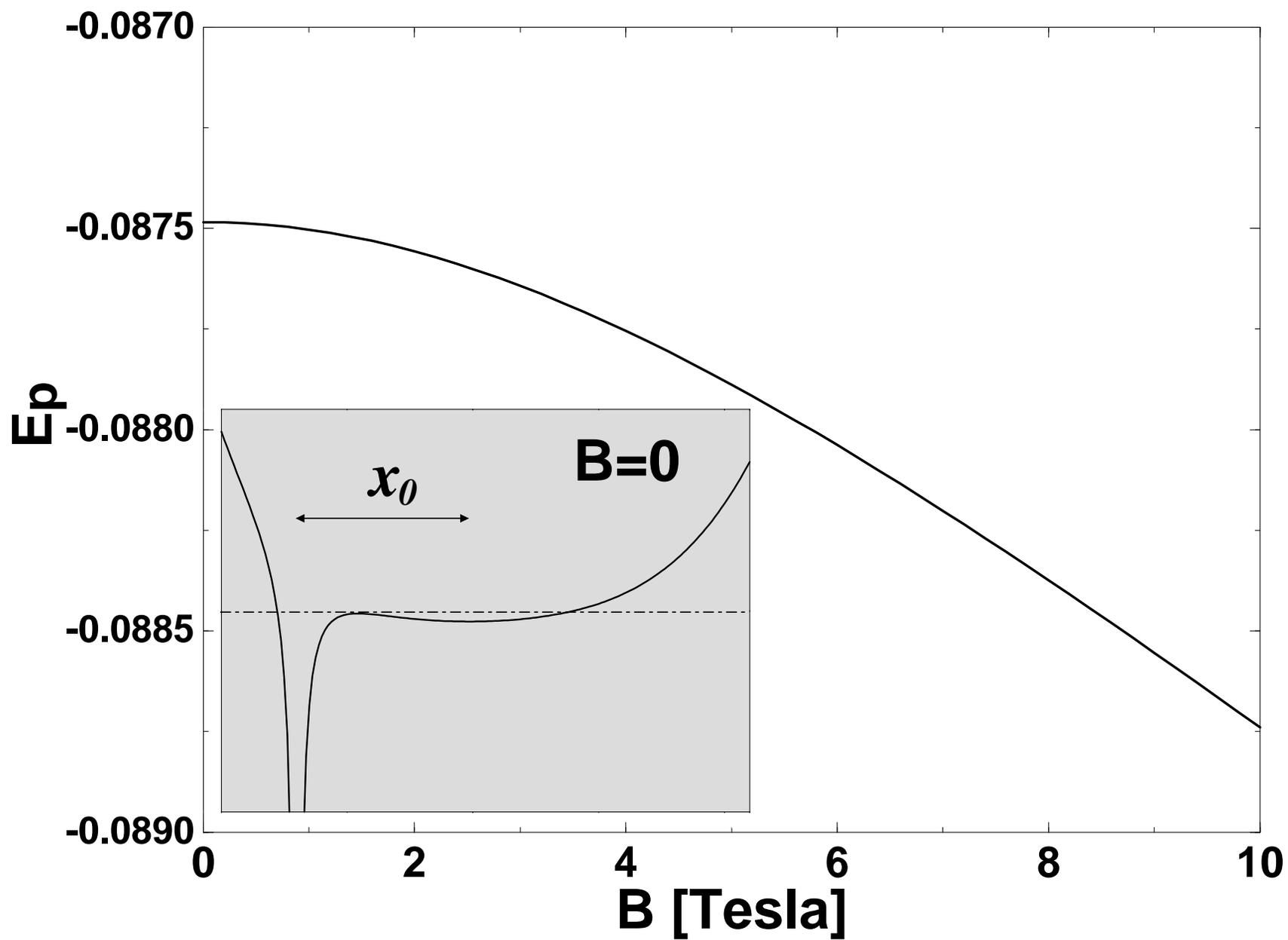

Figure 1: Yi Wan, Gerardo Ortiz and Philip Phillips

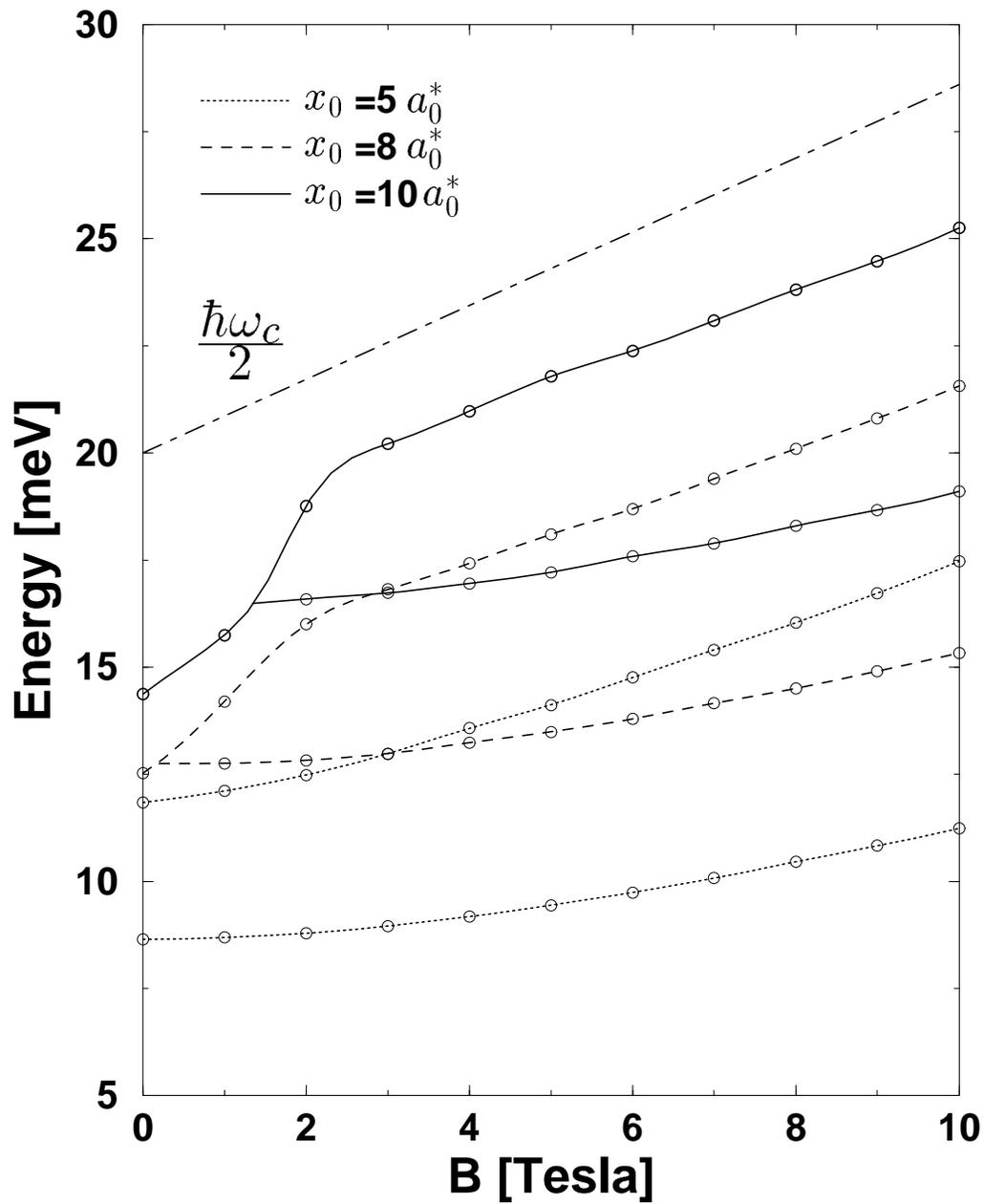

Figure 2: Yi Wan, Gerardo Ortiz and Philip Phillips

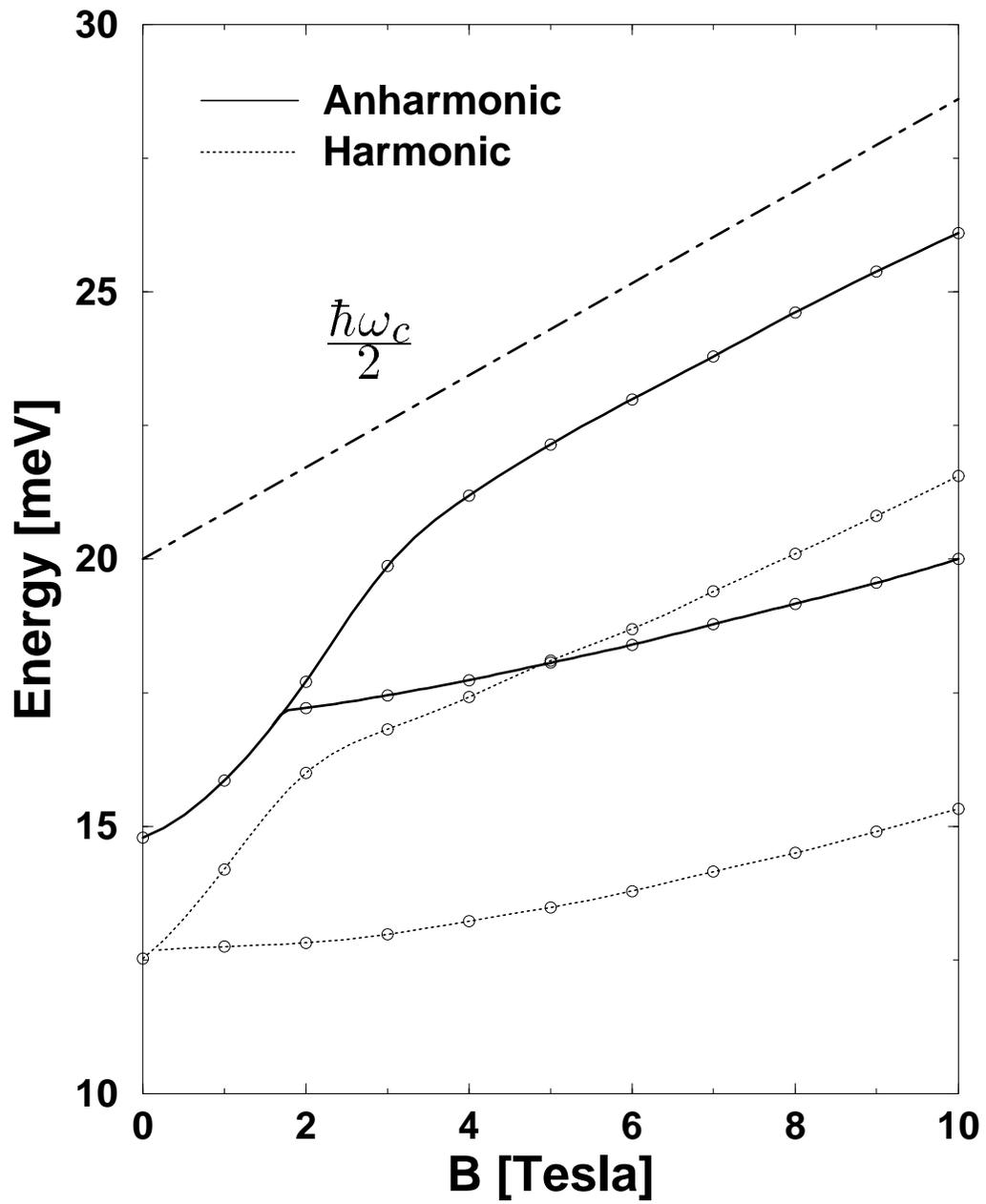

Figure 3: Yi Wan, Gerardo Ortiz and Philip Phillips

# Pair Tunneling in Semiconductor Quantum Dots

Yi Wan, Gerardo Ortiz, and Philip Phillips
*Loomis Laboratory of Physics*
*University of Illinois at Urbana-Champaign*
*1100 W.Green St., Urbana, IL, 61801-3080*

We propose here a model for the pair tunneling states observed by Ashoori and co-workers (Phys. Rev. Lett. **68**, 3088 (1992)) in GaAs quantum dots. We show that while GaAs is a weakly-polar semiconductor, coupling to optical phonons is sufficiently strong to mediate a negative-U pairing state. The physical potential in which the two electrons are bound can be composed of a Si impurity and a parabolic well that originates from the potential created by the $\delta$−dopants in the backing layer of the dot. Such a pair state breaks up at moderate magnetic field strengths ($\approx$ 2 T), as is seen experimentally, and is unstable when the confining radius of the dot is smaller than $\approx 400$Å.

Single electron capacitance spectroscopy[1] (SECS) has made it possible to detect single electron charging events in a semiconductor quantum dot that otherwise contains no excess electrons. In this regime, the Coulomb-blockade energy is negligible relative to the energy of local potential fluctuations. As the bias voltage is increased, electrons tunnel into local defect states on the quantum dot. Beyond a critical value of the bias voltage, sufficient charge has accumulated on the dot to form a uniform charging state. Transport becomes Coulomb-limited for all subsequent charging events[2,3]. Recently, Ashoori and co-workers[1] reported a series of measurements on a GaAs tunnel capacitor in the sub-Coulomb blockade regime. A tunnel capacitor is a layered structure consisting of a source lead, a tunnel junction, a laterally-confined GaAs quantum well, a $\delta$−doped backing layer of AlGaAs and a gate electrode[1]. The $\delta$-dopants in the backing layer ionize creating an excess of negative charge that repels the tunneling electrons from the backing layer. As a consequence, charge is restricted to flow back and forth between the source lead and the quantum well. The novel feature of these experiments is that single as well as two-electron tunneling events were observed[1] in quantum dots 1$\mu m$ in diameter. Direct confirmation of pair (or two electron)-tunneling is obtained from the evolution of the tunneling peaks in the presence of a magnetic field. The pair peaks, labeled $X$, $Y$, and $Z$ in Fig. 6b of Ref. [1b], were shifted to higher gate voltages ($V_G$) and were observed to split into two distinct tunneling peaks at critical values of the field, $B_c \approx 2$T. Above $B_c$, the spectroscopic signal now corresponds to two energetically-resolved tunneling events. The amplitude for each peak is consistent with that for single-electron tunneling. Hence, above $B_c$, each state carries a single electron, whereas the charge in the states $X$, $Y$, and $Z$ at zero-field must be $2e^-$. At a field of 10 T, the splitting between the two states for all three pair states is roughly 3 meV. The Zeeman energy at this field strength is 0.25 meV. Hence, breakup of the pair cannot be attributed to a simple Zeeman effect. Another curious feature is that the pair states were not observed[4] in defect-free quantum dots with a physical diameter of 3000Å or equivalently a lateral confining length of 400Å.

The question then arises, what is the source of the apparent pair-tunneling peaks in the capacitance spectrum of the 1$\mu m$ quantum dots? As GaAs is not a superconductor, it is surprising that a stable pair-tunneling state is observed at all. A theory of the elusive pair states in GaAs quantum dots must explain the following facts: 1) the stability of a $2e^-$ tunneling state at zero magnetic field, 2) a pair instability at a critical field $B = B_c$, 3) the emergence of two distinct tunneling peaks for $B > B_c$, and 4) the absence of pair-tunneling states in defect-free quantum dots with a confining diameter approximately $\leq 400$Å. We propose here a model which is capable of explaining each of these observations.

Although the magnetic field dependence might seem decoupled from the pairing mechanism, it turns out that an understanding of the field dependence is crucial to the formulation of a model for pair tunneling in quantum dots. Asymptotically, the energy of a parabolically-confined electronic state increases as $\hbar\omega_c/2$ in a magnetic field. Here $\omega_c$ is the cyclotron frequency. Inspection of the plots of $V_G$ vs $B$ in Fig. 6 of Ref. [1b] reveals that the energy of some of the tunneling states scales asymptotically as $\hbar\omega_c/2$, while others do not. States which exhibit the asymptotic $\hbar\omega_c/2$ slope originate from 2D harmonic confinement on a length scale of 300Å. Such confinement arises from the potential created by clustering of ionized Si-dopants in the AlGaAs backing layer[1,5]. Ashoori and co-workers have attributed the states whose energy is relatively insensitive to a magnetic field as arising from Si impurities that diffuse into the quantum well from the source lead[1]. In such hydrogenic impurities, the $1/r$ potential competes with the parabolic potential generated by the magnetic field to suppress the asymptotic $\hbar\omega_c/2$ slope. The effective Bohr radius of a Si impurity in GaAs is $a_0^* \approx 100$Å.

Consider now the pair states labeled $X$, $Y$, and $Z$. A consistent trend in these pair states is that above $B_c$,

the two electronic states evolve with fundamentally different slopes, indicating two distinct confining potentials. The upper state consistently has a slope close to $\hbar\omega_c/2$ whereas the slope of the lower state is indicative of that of a hydrogenic impurity. Consequently, the pair states appear to be hybrid states composed of a Si-impurity and a parabolic well formed from the $\delta$-dopants in the backing layer. A hybrid potential of this sort will give rise to two different length scales once the electron pair breaks up. Hence, such a potential will be the starting point for our analysis of the pair states. We point out that some experimental evidence exists in which both states of the pair evolve as $\hbar\omega_c/2$ above $B_c$, for example the data shown in Fig. 6a of Ref. [1b]. However, these states appear to be less stable and less reproducible than are the pair states of the $X$, $Y$, and $Z$ type because they transform into the latter upon thermal cycling of the sample[1]. We take then the $X$, $Y$, and $Z$ pairs to be the archetypal pair states and focus heretofore on explaining their origin.

Barring something truly exotic, the only tool available to mediate pair tunneling is coupling to optical phonons. GaAs is a weakly-polar semiconductor with an optical phonon energy of $\hbar\omega_{LO}=36.6$ meV and coupling constant of $\alpha = 0.08$. To determine whether phonons stabilize a pair state in a hybrid hydrogenic-parabolic potential on a quantum dot in an external magnetic field, we must solve the Hamiltonian $\tilde{H} = \hat{H}_0(1) + \hat{H}_0(2) + \eta/r_{12}$ for the two and one-particle ground-state energies, $\tilde{E}(2)$ and $\tilde{E}(1)$, respectively. In this expression[6], $\eta = e^2/\epsilon_0$ with $\epsilon_0$ (=12.53) the static dielectric constant for GaAs and $\hat{H}_0(i) = \hat{H}_1(i) + \hat{H}_{e-ph}$ is the sum of a 1-body Hamiltonian

$$\hat{H}_1(r) = \frac{\mathbf{P}^2}{2} - \frac{\eta}{r} + V_Q(z) + \frac{\omega_0^2}{2}\left[(x-x_0)^2 + y^2\right]$$
$$+ \frac{1}{2}\left(\frac{\omega_c}{2}\right)^2 (x^2 + y^2) \quad (1)$$

and the electron-phonon interaction of the Fröhlich type

$$\hat{H}_{e-ph} = \sum_k a_k^\dagger a_k$$
$$+ i\frac{g}{\sqrt{V}}\sum_k \frac{1}{k}\left(a_k^\dagger e^{-i\mathbf{k}\cdot\mathbf{r}} - a_k e^{i\mathbf{k}\cdot\mathbf{r}}\right), \quad (2)$$

where $(\mathbf{P}, \mathbf{r})$ are the electron momentum and position, $g = (\sqrt{2}\pi\alpha)^{1/2}$, $V$ is the crystal volume; the confining potential of the GaAs/AlGaAs quantum well of width $L$ is $V_Q(z) = V_0\,\theta(|z|-L/2)$, $\omega_0$ is the frequency of the oscillation in the parabolic well formed by the $\delta$-dopants and $a_k^\dagger$ creates an optical phonon with momentum $k$. In writing Eqs. (1-2) we have set $\hbar = m^* = \omega_{LO} = 1$. The coordinate $\mathbf{r}_0 = (x_0, 0, 0)$ determines the distance between the Coulomb and parabolic potentials. We have also chosen the symmetric gauge $\mathbf{A} = \frac{\omega_c}{2}(-y, x, 0)$; considering the range of magnetic fields involved and based on an analysis of $D^-$ centers[7,8] in GaAs, we ignore paramagnetic contributions to $\hat{H}_1$. Phonons will mediate a negative-U center if $\tilde{E}(2) - 2\tilde{E}(1) < 0$. Of course, this pair constraint necessitates solving the coupled electron-phonon problem exactly. The relevant quantity that must be calculated is the bi-polaron attractive interaction, $E_B$. However, as GaAs is in the weak electron-phonon coupling regime, $\alpha = .08 << 1$, it can be shown[9] that the result based on perturbation theory[10], $E_B = 2E_p$, is an excellent approximation, where $E_p$ is the polaron binding energy. Our approach will be to solve the electron problem separately using Quantum Monte Carlo (QMC) techniques. To this end, we redefine our working Hamiltonian to be $\hat{H} = \hat{H}_1(1) + \hat{H}_1(2) + \eta/r_{12}$. Let $E(2)$ and $E(1)$ be the exact ground state energies of $\hat{H}$ and $\hat{H}_1$, respectively. The pair-binding condition can now be recast as $E(2) + 2E_p - 2E(1) < 0$.

To obtain $E_p$, we need to solve for the phonon-enhancement in the binding energy of the ground state of $\hat{H}_0 = \hat{H}_1 + \hat{H}_{e-ph}$. For the 3D free electron problem, the polaron energy is simply the Fröhlich[11] result $E_p = -\alpha$ (in units of $\hbar\omega_{LO}$). Platzman[12] has shown that for an electron bound to a Coulomb potential, the polaron binding is enhanced in the strong coupling regime relative to the free polaron problem. We will show here that the polaron energy is further enhanced in our hybrid potential even in the weak coupling limit. This enhancement assists the formation of a pair state. To proceed, we use Feynman's approach[13] and obtain the effective Euclidean action in terms of the electron path, $\mathbf{r}_t$,

$$S_{eff} = -\frac{1}{2}\int dt\,\dot{\mathbf{r}}_t^2 + \frac{g^2}{4\pi}\int\int dt ds\,\frac{e^{-|t-s|}}{|\mathbf{r}_t - \mathbf{r}_s|} + \eta\int dt\,\frac{1}{r_t}$$
$$- \int dt\,V_Q(z_t) - \frac{1}{2}\left(\frac{\omega_c}{2}\right)^2\int dt\,(x_t^2 + y_t^2)$$
$$- \frac{\omega_0^2}{2}\int dt\,[(x_t - x_0)^2 + y_t^2] \quad (3)$$

by integrating out the phonon degrees of freedom. The presence of a retarded electron-phonon interaction, $1/r$, and quantum well potentials precludes an exact analytic solution to this problem. We are guaranteed, however, if we define a trial action $S_T$

$$S_T = S_1 - \frac{C}{2}\int\int dt ds\,|\mathbf{r}_t - \mathbf{r}_s|^2\,e^{-W|t-s|}$$
$$- \frac{K}{2}\int dt\,r_t^2 - \frac{K_z}{2}\int dt\,z_t^2 \quad (4)$$

that the total energy will satisfy the variational upper bound, $E = E_T - \lim_{\beta\to\infty}\beta^{-1}\langle S_{eff} - S_T\rangle \geq E_0$. We have defined $S_1$ to be the sum of the first and last two terms in the effective action and $E_T = -\lim_{\beta\to\infty}\beta^{-1}\ln\int \mathcal{D}\mathbf{r}_t \exp[S_T]$. The parameters, $C$, $W$, $K$, and $K_z$ are to be determined variationally. The polaron binding energy is given by $E_p = E - E(\alpha = C = 0)$. Computation of $E_p$ is compounded by the lack of central symmetry of the confining potential. Only for the case in which $x_0 = 0$ can the integrals be performed in closed form. It turns out that for a bound polaron, the explicit $x_0$ dependence for the separations considered here amounts only to a few percent correction to the polaron binding energy[9]. Hence, we will use only $E_p(x_0 = 0)$ solutions. We evaluated the path integrals using Laplace transform techniques to solve the saddle point equations. As the details of such calculations are rather lengthy, we only present the results for $E_p$. Results for the magnetic field dependence of the polaron binding energy are shown in Fig. 1. We find that $E_p$ has a weak field dependence. At 10 T, the increase in $|E_p|$ is 0.05 meV, which is five times smaller than the Zeeman energy (0.25 meV). Hence, $E_p$ is relatively insensitive to the magnetic field. At $B = 0$, the confining potential has enhanced the polaron binding energy from $E_p = -\alpha$ to $E_p = -1.1\alpha$. For GaAs, this corresponds to an enhancement of 0.3 meV or equivalently an enhancement of 0.6 meV for the bi-polaron binding energy. We will see that this enhancement plays a role in the pair-formation condition.

We now need to evaluate accurately the energies involving the electronic degrees of freedom. For the range of magnetic fields we are dealing with, the ground state of the 2-fermion system is a spin-singlet, which means that the configurational part of its wavefunction must be symmetric. We can solve exactly this boson problem by using projector QMC methods which are based on the property that in the asymptotic imaginary time ($\tau$) limit, the Euclidean evolution operator acting on a parent state $\Phi_T$ projects out the ground state $\Phi_0$: $\Phi_0 \propto \lim_{\tau \to \infty} \exp[-\tau(\widehat{H} - E_s)] \Phi_T$, where $E_s$ is a suitable trial energy which shifts the zero of the energy spectrum. We transform the time-dependent Schrödinger equation for $\Phi$ in Euclidean time $\tau$ to a master equation for the *importance-sampled* distribution $P(\mathcal{R}, \tau) = \Phi_T(\mathcal{R}) \Phi(\mathcal{R}, \tau)/\|\sqrt{\Phi_T \Phi}\|^2$,

$$\frac{\partial P(\mathcal{R}, \tau)}{\partial \tau} = \sum_{i=1}^{2} \nabla_i \cdot \left[\frac{1}{2}\nabla_i P(\mathcal{R}, \tau) - \mathbf{F}_i(\mathcal{R}) P(\mathcal{R}, \tau)\right] - (E_L(\mathcal{R}) - E_s) P(\mathcal{R}, \tau) , \quad (5)$$

and use stochastic random-walks in configuration space ($\mathcal{R} = (\mathbf{r}_1, \mathbf{r}_2)$) to solve this equation. $\mathbf{F}_i(\mathcal{R}) = \nabla_i \ln \Phi_T$ is the *drift velocity* and $E_L(\mathcal{R}) = \Phi_T^{-1} \widehat{H} \Phi_T$ is the local energy. The trial function

$$\Phi_T(\mathcal{R}) = \exp\left[\frac{r_{12}}{2 + ar_{12}}\right] [g_1(1)g_2(2) + g_1(2)g_2(1)]$$
$$\times \prod_i^2 \phi(z_i) \exp\left[-\frac{\alpha_1 r_i}{1 + \alpha_2 r_i} - \mu \rho_i^2\right] \quad (6)$$

is used to **guide** the random-walk. In Eq. 6 $g_1(i) = \exp[-br_i^2/(1 + cr_i)]$, $g_2(i) = \exp[-\nu((x - x_0)^2 + y^2)]$, $\phi(z) = \cos(\kappa z)$ or $\cos(\kappa L/2) \exp[\lambda(L/2 - |z|)]$ inside or outside the well, respectively, and with $a, b, c, \mu, \nu, \alpha_i, \kappa, \lambda$ variational parameters[9]. At sufficiently long times $P(\mathcal{R}, \tau \to \infty) \to \Phi_T(\mathcal{R}) \Phi_0(\mathcal{R})$ (up to a normalization constant), where $\Phi_0$ is the exact (nodeless) lowest energy state. In order to get this stationary distribution, $E_s$ must be adjusted to be equal to the ground state energy $E(2)$, given in turn by $E(2) = \lim_{\tau \to \infty} \langle E_L(\mathcal{R}) \rangle_{P(\mathcal{R}, \tau)}$. As long as $\Phi_T$ satisfies the right permutation symmetry, its functional form affects only the convergence and statistical fluctuations of $E(2)$. In a similar way we calculate $E(1)$.

We now combine our variational estimates for the polaron energy with our exact QMC calculations for $E(2)$ and $E(1)$ to solve the pair-binding condition, $\Delta_p = E(2) + 2E_p - 2E(1) < 0$. When pair-binding occurs, $\Delta_p < 0$ and both electrons reside in the same electronic state with energy $\epsilon = (E(2) + 4E_p)/2$ as measured in SECS. If $\Delta_p > 0$, two distinct states are occupied: a lower one with energy $\epsilon_1 = E(1) + E_p$ and a higher one with energy $\epsilon_2 = E(2) + 3E_p - E(1)$. Using the method outlined above, we have calculated $\Delta_p$ at three separations between the hydrogenic and parabolic wells: $x_0 = 5a_0^*$, $x_0 = 8a_0^*$, and $x_0 = 10a_0^*$. We have plotted in Fig. 2 the energy of two electrons on the quantum dot as a function of the magnetic field. The salient features of these graphs are as follows: 1) at $B = 0$, pair states form only when $x_0$ exceeds a critical value $\approx 8a_0^*$, 2) a magnetic field inhibits pair-binding at critical fields of $0.2T$ ($x_0 = 8a_0^*$) and $1.5T$ ($x_0 = 10a_0^*$) and 3) as in the experiments[1], two new states emerge above $B_c$; $\epsilon_1$ has a weak field dependence and $\epsilon_2$ scales asymptotically as $\hbar\omega_c/2$. The absence of pair-binding for $x_0 \leq 8a_0^*$ is a signature that the electronic wavefunctions are too localized for the electron-phonon attraction to outweigh the Coulomb repulsion. Similarly, increasing the magnetic field further confines the wavefunction of the electrons in the pair state. As a result, the Coulomb repulsions increase and at a critical value of the field, the polaron effect can no longer hold the pair together, resulting in pair-breaking.

Of course, the confining potential caused by the $\delta$-dopants is only approximately parabolic. Anharmonicities are undoubtedly present. We explored the role of an anharmonic correction of the form $\zeta ((x - x_0)^2 + y^2)^2$. This correction increases the barrier between the hydrogenic and parabolic potentials (see inset in Fig. 1) thus increasing the mean distance between the two electrons and consequently reducing the Coulomb repulsion. The results shown in Fig. 3 are for a separation of $x_0 = 8a_0^*$. As is evident, anharmonicity of this type increases pair-binding and pushes $B_c$ to a higher value of $\approx 2$T. In addition, the magnetic field dependence of $\epsilon_1$ and $\epsilon_2$ is as in the harmonic case. We conclude then that the experimental observations of pair-binding can be explained by a model in which optical phonons bind two electrons in a Si impurity and a roughly parabolic well (created by the $\delta$-dopants) that are sufficiently far apart[14]. The critical separation even when anharmonic corrections are included seems to be $\approx 7a_0^* = 700$Å. We would predict then an absence of pair-binding in quantum dots that are either clean (no Si) or laterally smaller than $\approx 7a_0^*$ in confining diameter, consistent with experiments reported in Ref. [1b]. We close by pointing out that the polaron binding considered here increases monotonically as the GaAs quantum well width diminishes, and in the strict 2-dimensional limit[10,9], $E_p = -\pi\alpha/2$. Such an enhancement is sufficient to cause pair-binding in samples where **both** lateral confining radii and quantum well widths are reduced. Hence, a critical test of the model reported here would be a detailed sample-size dependent study of the stability of pair states in GaAs quantum dots.

## ACKNOWLEDGMENTS

We thank Ray Ashoori for discussions illuminating the details of the experimental measurements. This work is supported in part by the NSF grants No. DMR94-96134 and DMR-91-17822.

FIG. 1. Polaron binding energy $E_p$ (in units of $\hbar\omega_{LO}$) as a function of external magnetic field $B$. The inset is a schematic plot of the ($y = z = 0$) electron confining potential responsible for pair formation, $x_0$ being the distance between hydrogenic and parabolic components.

FIG. 2. Electron energies as a function of magnetic field $B$. For each pair-potential separation $x_0$, the lower branch is $\epsilon_1 - E_p$ while the upper one corresponds to $\epsilon_2 - E_p$. The open circles represent the QMC results whose statistical error bars are 0.06 meV (smaller than the size of the circles). The curves are spline interpolations to the QMC data. The dashed-dotted line represents the field dependence of the lowest Landau level $\hbar\omega_c/2$.

FIG. 3. Effect of anharmonicities on the electron energies at pair-potential separation $x_0 = 8a_0^*$. Note anharmonic corrections shift $B_c$ to higher values. The meaning of the curves and open circles is the same as in Fig.2.